# Capillary Action in a Crack on the Surface of Asteroids with an Application to 433 Eros

Yu Jiang[1, 2], Hexi Baoyin[2]

1. State Key Laboratory of Astronautic Dynamics, Xi'an Satellite Control Center, Xi'an 710043, China

2. School of Aerospace Engineering, Tsinghua University, Beijing 100084, China

Y. Jiang (✉) e-mail: jiangyu_xian_china@163.com (corresponding author)

**Abstract**. Some asteroids contain water ice, and a space mission landing on an asteroid may take liquid to the surface of the asteroid. Gas pressure is very weak on the surface of asteroids. Here we consider the capillary action in a crack on the surface of irregular asteroids. The crack is modelled as a capillary which has a fixed radius. An asteroid's irregular gravitational potential influences the height of the liquid in the capillary. The height of the liquid in the capillary on the surface of such asteroids is derived from the asteroid's irregular gravitational potential. Capillary mechanisms are expected to produce an inhomogeneaous distribution of emergent liquid on the surface. This result is applied to asteroid 433 Eros, which has an irregular, elongated, and concave shape. Two cases are considered: 1) we calculate the height of the liquid in the capillary when the direction of the capillary is perpendicular to the local surface of the asteroid; 2) we calculate the height of the liquid in the capillary when the direction of the capillary is parallel to the vector from the center of mass to the surface position. The projected height in the capillary on the local surface of the asteroid seems to depend on the assumed direction of the capillary.

## 1. Introduction

Previous studies have claimed that Earth's water originated from asteroids (Morbidelli et al. 2000; Mottl et al. 2007; Campins et al. 2010). Kanno et al. (2003) analyzed the wavelength of infrared spectra and confirmed the presence of water ice on a D type asteroid. A meteoroid may be released from an asteroid following a collision and thus bring material to Earth (Treiman et al. 2004; Vereš et al. 2008; Ray and Misra 2014;

Patil et al. 2015). A meteoroid may also be released from Mars and bring water to Earth (Mitton 1992). Treiman et al. (2004) studied the Serra de Magé eucrite meteorite, which presumably came from asteroid 4 Vesta, and found quartz in the meteorite; they concluded that the quartz was deposited by liquid water, and the water probably came from outside 4 Vesta. Campins et al. (2010) reported that there exist water ice on the surface of asteroid 24 Themis, and that the water ice has a widespread distribution. Comets also contain water ice. Sunshine et al. (2006) detected solid water ice on the surface of comet 9P/Tempel and pointed out that the surface deposits are loose aggregates. Taylor (2015) reported that water exists in the Eucrite meteorites which came from asteroid 4 Vesta.

We focus here on the height of the liquid surface water on the surface of an asteroid, which is related to the surface equilibrium and surface motions on the asteroid. On the surface of asteroids in the inner Solar system, liquid water can exist (Cohen and Coker 1999; Yurimoto et al. 2014). Besides, a very transient presence of material in the liquid phase can exist in the active areas of cometary nuclei by sloar heating (Miles and Faillace 2001). Previous works have discussed the dynamics of surface equilibria on a rotating ellipsoid (Guibout and Scheeres 2003), the hopping on a flat surface of a rotating ellipsoid (Bellerose and Scheeres 2008; Bellerose et al. 2009), the contact motion and impact on the surface of asteroid 25143 Itokawa (Tardivel and Scheeres 2014), as well as the topological classification of equilibria in the potential of asteroids (Jiang et al. 2014). Guibout and Scheeres (2003) found that the stability of surface equilibria on an ellipsoid ties to the shape of the ellipsoid.

Bellerose and Scheeres (2008) discussed the dynamical behavior around the stable and unstable surface equilibria on a rotating ellipsoid to model the motion on an asteroid's surface. Jiang et al (2014) used an accurate asteroidal shape model and discussed equilibria and motion around equilibria in the potential of four asteroids: (216) Kleopatra, (1620) Geographos, (4769) Castalia, and (6489) Golevka. Jiang (2015) finds that the equilibrium stability and the stability of periodic orbits around equilibria have some corresponding relationships. However, the height that surface water reaches on an asteroid depends on the irregular shape and gravitational potential; different surfaces produce different heights, which may lead to different friction factors and also affect the stability of equilibria.

We model a crack on the asteroids as a capillary. We study the height that a liquid can reach within this capillary that is located on the surface of irregular asteroids. The height of liquid in the capillary is constant when the position of the capillary varies over the surface of a spherical-shaped body; however, the height is time-variant when the position of the capillary varies over the surface of an irregular-shaped asteroid. The results can be applied to two areas. First, we can study the water ice distribution on asteroids; different heights of the liquid in the capillary may produce different distributions of water ice on the surface (Campins et al. 2010). Second, the height a liquid reaches can affect the electrostatic and rotational ejection of gas and dust grains (Oberc 1997) on the surface of minor celestial bodies. Under the effect of the solar radiation pressure, the ejection will form a mini-fountain on the surface; the change in height of the liquid in the capillary causes the height and radius of the fountain envelope to vary (Oberc 1997).

The gravitational field of asteroids influences the height a liquid can reach in a capillary; this is the case for a single asteroid, binary asteroids, and multiple asteroid systems. We present the height a liquid in a capillary can reach on the surface of an asteroid in a multiple asteroid system. Asteroid 433 Eros is taken as an example because the surface shape is irregular, elongated, and concave. The liquid's height depends on the direction of the capillary; we calculated two cases, in the first case, the capillary's direction is perpendicular to the asteroid's local surface; in the second case, the capillary's direction is parallel to the line segment from the asteroid's center of mass to the surface position. The results show that a fluid can be brought to the surface from the interior by capillary mechanisms. The process is inhomogeneous, and it is more likely that certain regions of the surface can be much more efficient than others, due to the interplay of shape, gravitational field and possible density of cracks.

**2. The Height of a Liquid in the Capillary on the Surface of Asteroids**

Let us consider a crack on the surface of an asteroid. As we stated in the previous section, we use a capillary which has a fixed radius to model the crack. Denote $r_c$ as the radius of the capillary. The contact angle $\theta$ is defined as the angle between the liquid's surface and the outline of the solid's contact surface. The radius of curvature of the liquid surface is denoted as $R$, where $R$ is the radius of a circle which best fits the liquid surface's normal section, $R = -\dfrac{r_c}{\cos\theta}$. Let $\gamma$ be the surface tension, then according to the Jurin's rule, the liquid height in the capillary is $h = \frac{2\gamma \cos\theta}{\rho g_a r_c}$, where $\rho$ is the liquid density and $g_a$ is the gravitational acceleration on the surface of the asteroid. Let $\alpha$ be the angle between the capillary and the direction of the gravitational force, and $l$ be the length of liquid in the capillary, then $l = \frac{h}{\sin\alpha}$.

The gravitational field of an asteroid can be computed by assuming a shape approximated by means of a polyhedron model using observational data. The test point has the Cartesian coordinates $(x, y, z)$, the position of the differential mass dm is $(\xi, \eta, \varsigma)$, then the vector from the test point to the differential mass is $\mathbf{r} = (\xi - x, \eta - y, \varsigma - z) = (\Delta x, \Delta y, \Delta z)$. Using the polyhedron model, the asteroid's gravitational potential (Werner 1994; Werner and Scheeres 1997) can be expressed by Eq. (1):

$$U = G \iiint_{Body} \frac{1}{r} dm = \frac{1}{2} G\sigma \sum_{e \in edges} \mathbf{r}_e \cdot \mathbf{E}_e \cdot \mathbf{r}_e \cdot L_e - \frac{1}{2} G\sigma \sum_{f \in faces} \mathbf{r}_f \cdot \mathbf{F}_f \cdot \mathbf{r}_f \cdot \omega_f . \quad (1)$$

In addition, the gravitational force (Werner and Scheeres 1997) is given by Eq. (2):

$$\nabla U = -G\sigma \sum_{e \in edges} \mathbf{E}_e \cdot \mathbf{r}_e \cdot L_e + G\sigma \sum_{f \in faces} \mathbf{F}_f \cdot \mathbf{r}_f \cdot \omega_f , \quad (2)$$

and the Hessian matrix of the gravitational potential is given by Eq. (3):

$$\nabla(\nabla U) = G\sigma \sum_{e \in edges} \mathbf{E}_e \cdot L_e - G\sigma \sum_{f \in faces} \mathbf{F}_f \cdot \omega_f , \quad (3)$$

where G = 6.67 $\times 10^{-11}$ m$^3$kg$^{-1}$s$^{-2}$ is the gravitational constant, $r$ is the norm of $\mathbf{r}$, $\sigma$ is the bulk density of the asteroid, $\nabla$ is the gradient operator; $\mathbf{r}_e$ and are $\mathbf{r}_f$ are body-fixed vectors, $\mathbf{r}_e$ is from the test point to some fixed point on the polyhedron's edge, while $\mathbf{r}_f$ is from the test point to the point on the polyhedron's surface; $\mathbf{E}_e$ and $\mathbf{F}_f$ are edge- and face- dyads, respectively; $L_e$ represents the integration factor between the test point and the polyhedron's edge, $L_e = \ln\frac{a+b+e}{a+b-e}$, $a$ and $b$ are distances between the test point and the edge's two ends, $e$ is the edge length, $\omega_f$ represents the signed solid angle subtended by the triangle region when viewed from the test point.

Let **ω** be the rotational velocity of the asteroid, and $\omega$ be the norm of the vector **ω**, then the body-fixed frame is defined through $\mathbf{\omega} = \omega\mathbf{e}_z$. Then the effective potential $V$ and its gradient (Jiang and Baoyin 2014; Jiang 2015) can be given by Eqs. (4) and (5):

$$V = U - M\frac{\omega^2}{2}(x^2 + y^2), \text{ and} \tag{4}$$

$$\begin{cases} \dfrac{\partial V(\mathbf{r})}{\partial x} = -M\omega^2 x + \dfrac{\partial U(\mathbf{r})}{\partial x} \\ \dfrac{\partial V(\mathbf{r})}{\partial y} = -M\omega^2 y + \dfrac{\partial U(\mathbf{r})}{\partial y} \\ \dfrac{\partial V(\mathbf{r})}{\partial z} = \dfrac{\partial U(\mathbf{r})}{\partial z} \end{cases}. \tag{5}$$

where $M$ is the asteroid's mass. Then the gravitational acceleration of the liquid on the surface of the asteroid is given by Eq. (6):

$$g_a = \nabla V. \tag{6}$$

Assume the capillary's direction is parallel to the line segment from the asteroid's center of mass to the surface position. Substituting Eq. (6) into the expression for the height of the liquid yields Eq. (7):

$$h = \frac{2m\gamma G\sigma\cos\theta}{\rho r_c \nabla V}, \tag{7}$$

where $m$ is the liquid mass.

If the asteroid is a contact binary asteroid or a contact multiple asteroid, the gravitational potential can be written as Eq. (8):

$$U = -\sum_{i=1}^{n-1}\sum_{j=i+1}^{n}\int_{\beta_i}\int_{\beta_j}\frac{G\sigma(\mathbf{D}_i)\sigma(\mathbf{D}_j)dV(\mathbf{D}_j)dV(\mathbf{D}_i)}{\|\mathbf{A}_i\mathbf{D}_i - \mathbf{A}_j\mathbf{D}_j + \mathbf{r}_i - \mathbf{r}_j\|}, \qquad (8)$$

where $\beta_i$ is the $i$th contact body, subscript $i$ represents the parameter of $i$th contact body, $n$ is the number of bodies, $\mathbf{A}_i$ is the transformation matrix of $\beta_i$'s principal reference frame relative to the inertial space, $\mathbf{D}_i$ is the vector of mass element $dM(\mathbf{D}_i) = \sigma(\mathbf{D}_i)dV(\mathbf{D}_i)$ relative to the body-fixed frame of $\beta_i$, $dV(\mathbf{D}_i)$ is the volume element in $\beta_i$, $\sigma(\mathbf{D}_i)$ is the bulk density of $dV(\mathbf{D}_i)$, and $\mathbf{r}_i$ is the position vector of the barycentre of $\beta_i$ relative to the position vector of the barycentre of the asteroid. The asteroid's kinetic energy can be expressed by Eq. (9):

$$T = \frac{1}{2}\sum_{i=1}^{n}\left(M_i\|\dot{\mathbf{r}}_i\|^2 + \langle\boldsymbol{\omega}, \mathbf{I}_i\boldsymbol{\omega}\rangle\right), \qquad (9)$$

where $M_i$ is the mass of the $i$th contact body $\beta_i$, $\mathbf{I}_i$ is the moment of inertia of $\beta_i$. The force acting on the body $\beta_i$ is then given by Eq. (10):

$$\mathbf{f}_i = -G\sum_{j=1, j\neq i}^{n}\int_{\beta_i}\int_{\beta_j}\frac{(\mathbf{A}_i\mathbf{D}_i - \mathbf{A}_j\mathbf{D}_j + \mathbf{r}_i - \mathbf{r}_j)}{\|\mathbf{A}_i\mathbf{D}_i - \mathbf{A}_j\mathbf{D}_j + \mathbf{r}_i - \mathbf{r}_j\|^3}dM(\mathbf{D}_j)dM(\mathbf{D}_i). \qquad (10)$$

Then the effective potential $V$ for the liquid on the surface of the body $\beta_i$ becomes Eq. (11):

$$V_i = \sum_{\substack{j=1\\j\neq i}}^{n}\int_{\beta_i}\int_{\beta_j}\frac{G\sigma(\mathbf{D}_i)\sigma(\mathbf{D}_j)dV(\mathbf{D}_j)dV(\mathbf{D}_i)}{\|\mathbf{A}_i\mathbf{D}_i - \mathbf{A}_j\mathbf{D}_j + \mathbf{r}_i - \mathbf{r}_j\|} - \frac{1}{2}\langle\boldsymbol{\omega}, \mathbf{I}_i\boldsymbol{\omega}\rangle$$
$$+ \frac{1}{2}\int_{\beta_i}\int_{\beta_i}\frac{G\sigma(\mathbf{D}_i)\sigma(\mathbf{D}_i')dV(\mathbf{D}_i')dV(\mathbf{D}_i)}{\|\mathbf{A}_i\mathbf{D}_i - \mathbf{A}_i\mathbf{D}_i' + \mathbf{r}_i - \mathbf{r}_i'\|}. \qquad (11)$$

Let the capillary's direction be parallel to the line segment that stretches from the asteroid's center of mass to the surface position. Then the height of the liquid is given by Eq. (12):

$$h = \frac{2m\gamma G\sigma \cos\theta}{\rho r_c \nabla V_i}. \tag{12}$$

From Eqs. (11) and (12), we know that for an isolated asteroid, the height in a capillary that is fixed on the surface is constant. If the position on the asteroidal surface varies, the height will also vary. Whereas, in the case of the binary asteroids or multiple asteroid systems, the height in the capillary on a fixed surface position on one of the asteroids in the asteroidal system is time-varying.

## 3. Gravitational Effective Potential and Gravitational Slope

We apply the results to asteroid 433 Eros. Researchers have already developed shape models for a few asteroids, including 216 Kleopatra (Neese 2004), 951 Gaspra (Stooke 2002), and 2063 Bacchus (Neese 2004). Only asteroid 433 Eros has an irregular, elongated, mostly convex shape, but also including the presence of some concavities. Besides, the density, shapes and gravitational model for asteroid 433 Eros is accurate enough. So we use asteroid 433 Eros to compute the surface height of the liquid.

The physical model of asteroid 433 Eros has been calculated using radar observations from Gaskell (2008) with the polyhedral model (Werner 1994; Werner and Scheeres 1997). The estimated bulk density of asteroid 433 Eros is 2.67 g·cm$^{-3}$ (Miller et al. 2002), the moment of inertia is $17.09 \times 71.79 \times 74.49 \text{ km}^2$ (Miller et al. 2002), its rotational period is 5.27025547 h (Petit et al. 2014) and has overall dimensions of $(36 \times 15 \times 13)$ km (Petit et al. 2014). Fig. 1 shows the contour line of the effective potential for asteroid 433 Eros in the xy plane. The irregular shape of 433 Eros influences the contour line of the effective potential. From Fig. 1, one can see that the value of effective potential in the xy plane has five equilibrium points; these five equilibrium points are extreme points in the xy plane, not extreme points in the whole gravitational field of 433 Eros. The value of the effective potential near the surface of 433 Eros is about 50~60 $m^2s^{-2}$. In the concave area on the surface, the value of the effective potential is about 60 $m^2s^{-2}$; while in the convex area on the surface, the value of the effective potential is about 50 $m^2s^{-2}$.

Consider a capillary located on the surface of asteroid 433 Eros. We can calculate the height distributions of the liquid in the capillary. Figure 1 shows the contour line of the effective potential for asteroid 433 Eros in the xy-plane, the unit of the effective potential is 1.0 $m^2s^{-2}$. From Figure 1, one sees that the value of the effective potential changes greatly near the concave area; the concavity extends along x- and x+. So we plot figures viewed from the -x axis to see the obvious changes in the height distributions. The surface data from Stooke (2002), Neese (2004), and Gaskell (2008) can generate a polyhedral model (Werner 1994; Werner and Scheeres 1997) of asteroids. However, at the edge of the polyhedral model, the surface is not smooth and leads to a large error. To study the effect of the surface on the height more accurately, we use the Nagata patch interpolation (Tong and Kim 2009; Neto et al. 2013) to calculate the contact surface of the capillary and the asteroid.

The Nagata edge's interpolation and the curve's derivative (Neto et al. 2013) are given by Eq. (13):

$$\begin{cases} \mathbf{r}(\xi) = \mathbf{r}_0 + (\mathbf{r}_1 - \mathbf{r}_0 - \mathbf{c})\xi + \mathbf{c}\xi^2 \\ \mathbf{r}_\xi(\xi) = (\mathbf{r}_1 - \mathbf{r}_0) + (2\xi - 1)\mathbf{c} \end{cases}, \quad (13)$$

where $\xi$ is the local coordinate on the surface which satisfies $0 \leq \xi \leq 1$, $\mathbf{r}_0$ and $\mathbf{r}_1$ are the locations of the vertex, $\mathbf{c}$ is a coefficient. A triangle on the asteroid's surface has three vertices $v_1$, $v_2$, and $v_3$; the positions of these three vertices are denoted as $\mathbf{r}_{00}$, $\mathbf{r}_{10}$, and $\mathbf{r}_{11}$, respectively; the normal vectors at these vertices are denoted as $\mathbf{n}_{00}$, $\mathbf{n}_{10}$, and $\mathbf{n}_{11}$, respectively. Then for the triangular patch (Neto et al. 2013) of the asteroid's polyhedral model, the interpolated surface can be calculated using Eq. (14):

$$\mathbf{r}(\eta, \varsigma) = \mathbf{c}_{00} + \mathbf{c}_{10}\eta + \mathbf{c}_{01}\varsigma + \mathbf{c}_{11}\eta\varsigma + \mathbf{c}_{20}\eta^2 + \mathbf{c}_{02}\varsigma^2, \quad (14)$$

where $\mathbf{r}(\eta, \varsigma)$ represents any point's position vector on the triangular patch. $\eta$ and $\varsigma$ are local coordinates on the triangular patch which satisfy $0 \leq \eta \leq 1$ and $0 \leq \varsigma \leq 1$; $\mathbf{c}_{00}, \mathbf{c}_{10}, \mathbf{c}_{01}, \mathbf{c}_{11}, \mathbf{c}_{20},$ and $\mathbf{c}_{02}$ are coefficient vectors which satisfy the following conditions shown by Eq. (15):

$$\begin{cases} \mathbf{c}_{00} = \mathbf{r}_{00} \\ \mathbf{c}_{10} = \mathbf{r}_{10} - \mathbf{r}_{00} - \mathbf{c}_1 \\ \mathbf{c}_{01} = \mathbf{r}_{11} - \mathbf{r}_{10} + \mathbf{c}_1 - \mathbf{c}_3 \\ \mathbf{c}_{11} = \mathbf{c}_3 - \mathbf{c}_1 - \mathbf{c}_2 \\ \mathbf{c}_{20} = \mathbf{c}_1 \\ \mathbf{c}_{02} = \mathbf{c}_2 \end{cases}. \quad (15)$$

The normal vector (Bastl et al. 2008; Neto et al. 2013) at any point on the triangular patch can be calculated by the cross product of the two vectors given by Eq. (16):

$$\begin{cases} \mathbf{r}_\eta(\eta,\varsigma) = \dfrac{\partial \mathbf{r}}{\partial \eta} = \mathbf{c}_{10} + \mathbf{c}_{11}\varsigma + 2\mathbf{c}_{20}\eta \\ \mathbf{r}_\varsigma(\eta,\varsigma) = \dfrac{\partial \mathbf{r}}{\partial \varsigma} = \mathbf{c}_{01} + \mathbf{c}_{11}\eta + 2\mathbf{c}_{02}\varsigma \end{cases}. \tag{16}$$

If the capillary's direction is perpendicular to the asteroid's local surface, the height in the capillary can be calculated using Eq. (17):

$$\begin{cases} \mathbf{r}_{\eta\varsigma} = \mathbf{r}_\eta(\eta,\varsigma) \times \mathbf{r}_\varsigma(\eta,\varsigma) \\ h = \dfrac{2m\gamma G\sigma \cos\theta}{\rho r_c \nabla V_i} \cdot \dfrac{\mathbf{f}_i \cdot \mathbf{r}_{\eta\varsigma}}{|\mathbf{f}_i||\mathbf{r}_{\eta\varsigma}|} \end{cases}. \tag{17}$$

To calculate the heights of the liquid in the capillary on the surface of asteroid 433 Eros, we take for initial values: the radius of the capillary is $r_c = 0.1 \text{cm}$, the contact angle $\theta = 3.0°$, and the liquid density $\rho = 1.0 \text{g} \cdot \text{cm}^{-3}$. When the temperature is 20° C, the mercury's surface tension is $485 \times 10^{-3} \text{N} \cdot \text{m}^{-1}$, the acetic acid's surface tension is $27.6 \times 10^{-3} \text{N} \cdot \text{m}^{-1}$, and the water's surface tension is $72.75 \times 10^{-3} \text{N} \cdot \text{m}^{-1}$. We now let the surface tension for our example be $\gamma = 70.0 \times 10^{-3} \text{N} \cdot \text{m}^{-1}$. We now calculate the height of the liquid in the capillary when the direction of the capillary is perpendicular to the local surface and when it is parallel to the vector from the center of mass to the surface position.

Fig. 2 shows the heights of the liquid in the capillary when the direction of the capillary is perpendicular to the local surface of the asteroid 433 Eros. It can be seen that there exist several local maximum points and local minimum points for the heights. At the concave area, there is a global maximum point. Asteroid 433 Eros' shape is elongated; the heights at both ends are low. Near the concave area, there are two convex areas; at these two convex areas, the heights are lower than those at the concave area.

Fig. 3 shows the heights of the liquid in a capillary when the direction of the capillary is parallel to the vector from the center of mass to the surface position of asteroid 433 Eros. It can be seen that there exist several local maximum points and local minimum points for heights. At the concave area, there is a global maximum point. Similar to the case of a capillary perpendicular to the local surface shown in Fig. 2, the heights at both ends are low, and at the two convex areas near the concave area the heights are lower than those at the concave area.

Fig. 4 shows the projected heights of the liquid in the capillary when the direction of the capillary is parallel to the vector from the center of mass to the surface position. We can see that there exist several local maximum points and local minimum points for the projected lengths. Unlike Fig. 3 and Fig. 4, the global maximum point is not at the concave area.

Compare Fig. 4 with Fig. 2 and Fig. 3. One can see that the distribution of the projected heights is quite different from the two cases that show the distribution of the heights. The projected height depends on the direction normal to the surface. The liquid heights in the two cases are very low, which depend on the radius of the capillary, the contact angle, the liquid density, as well as the distribution of the irregular gravitation of the asteroid. If the radius of the capillary is smaller, for instance, let $r_c = 0.01 \text{cm}$, then the liquid heights become 10 times greater than before. The number of capillary paths depends on the structure of the asteroid. From the result, although we cannot positively prove or rule out that there are significant or negligible amounts of ice on the asteroid surfaces, we know that that the liquid heights on different areas of the asteroid's surface must be different due to the effect of an irregular gravitational field on the sublimation of the ices.

## 4. Conclusions

The capillary action in a crack which is on the surface of irregular asteroids has been investigated. The crack is modelled as a capillary. The capillary is assumed to have a fixed radius to reduce the required CPU time needed for the computations. The formulas for the heights of the liquid in the capillary have been derived. The height depends on the asteroid's irregular gravitational potential. We applied the formulas to asteroid 433 Eros. We considered two different orientations of the capillary, one with the direction of the capillary perpendicular to the local surface of the asteroid 433 Eros, the other with the direction of the capillary parallel to the vector from the center of mass to the surface position of asteroid 433 Eros. For both cases, there exist several local maximum points and local minimum points for the heights of the liquid, and the heights at both ends are low. Near the concave area of 433 Eros, there are two convex areas; at these two convex areas, the heights are lower than that at the concave area.

For the case when the direction of the capillary is parallel to the vector from the center of mass to the surface position, the projected height of the liquid is quite different from the above two cases. The global maximum point is not at the concave area. The projected height on the asteroid surface depends on the direction normal to the surface. The conclusions are also suitable for comets. The rotational periods of comets, in general, are larger than asteroids; which makes the gravitational slope are different. Since the liquid height depends on the gravitational slope, the distribution of liquid height on the surface of comets is different from asteroids. The height of the fluid in a given capillary is constant on the surface of a single object, whereas it is periodically variable in the presence of one companion in the binary or multiple systems.

**Acknowledgements**

This research was supported by the National Natural Science Foundation of China (No. 11372150), the State Key Laboratory of Astronautic Dynamics Foundation (No. 2015ADL-0202) and the National Basic Research Program of China (973 Program, 2012CB720000).

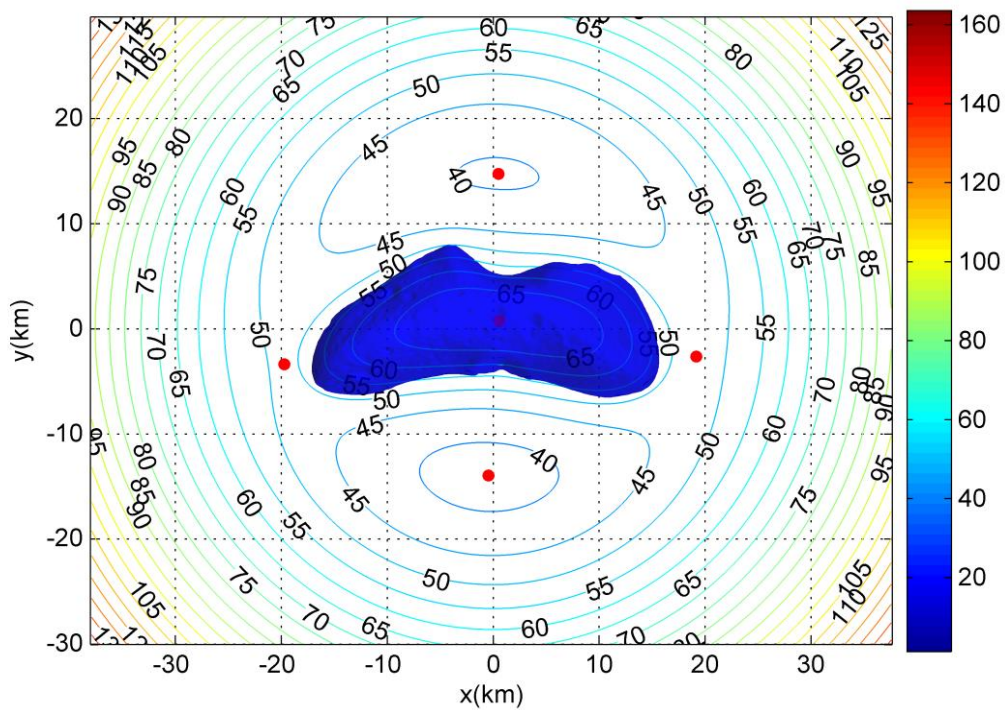

Figure 1. The contour line of the effective potential for asteroid 433 Eros in the xy-plane.

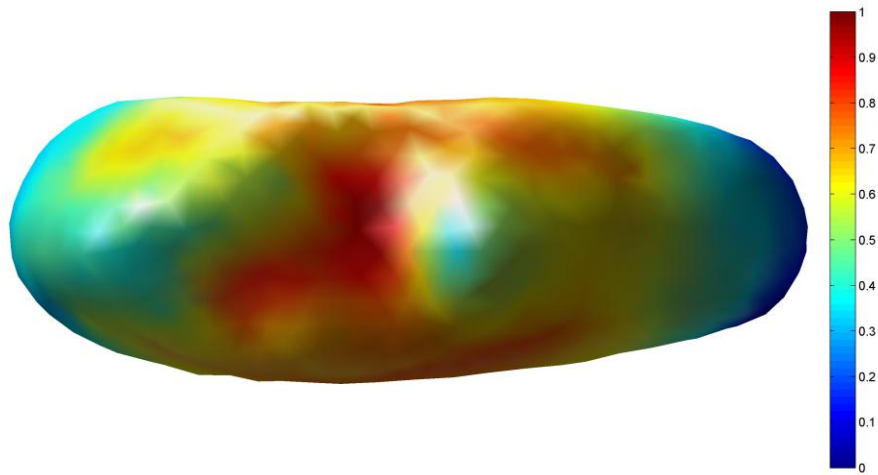

Figure 2. The liquid height in the capillary (view from +x axis), the direction of the capillary is perpendicular to the local surface of the asteroid 433 Eros, the unit of liquid height is 0.24737427667173 cm.

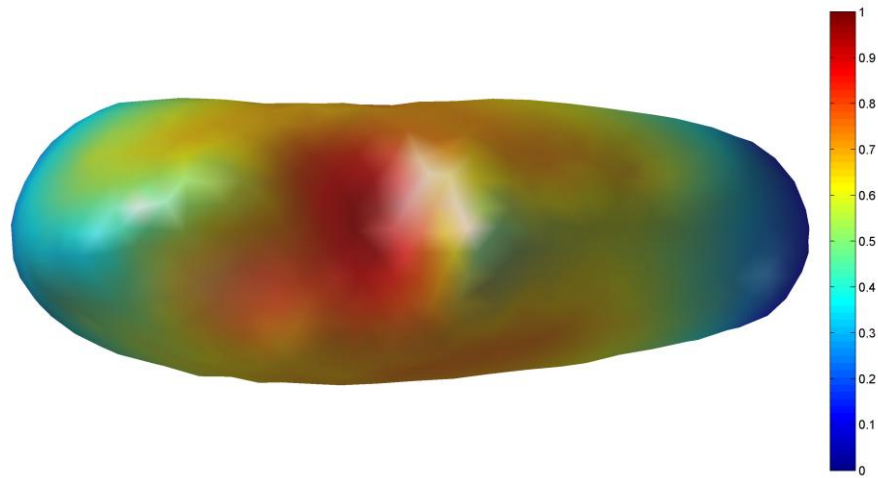

Figure 3. The liquid height in the capillary(view from +x axis), the direction of the capillary is parallel to the vector from the center of mass to the surface position of asteroid 433 Eros, The unit of liquid height is 0.23498122454119 cm.

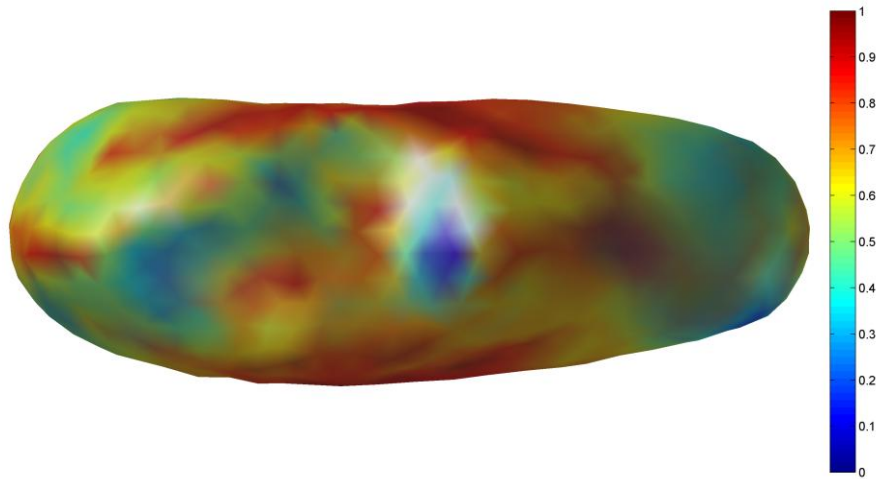

Figure 4. The length of projection of liquid height in the capillary (view from +x axis), the direction of the capillary is parallel to the vector from the center of mass to the surface position, the unit of liquid height is 0.12138046222674 cm.